\documentstyle[12pt]{article}
\input{epsfig.sty}
\newcommand{\beq}{\begin{eqnarray}}
\newcommand{\eeq}{\end{eqnarray}}
\begin{document}
\title{$\Psi(2S)$, $\Upsilon(3S)$ Suppression in p-Pb, Pb-Pb Collisions and 
Mixed Hybrid Theory}
\author{Leonard S. Kisslinger\\
Department of Physics, Carnegie Mellon University, Pittsburgh, PA 15213}
\date{}
\maketitle
\begin{abstract} We use our mixed hybrid model for the $\Psi(2S)$ state
to estimate $\Psi(2S)$ to $J/\Psi(1S)$ suppression in p-Pb collisions, and 
the $\Upsilon(3S)$ state to estimate $\Upsilon(3S)$ to $\Upsilon(1S)$ 
suppression in Pb-Pb collisions, and compare to recent experimental 
measurements.
\end{abstract}
\vspace{1mm}

\section{Introduction}
  The production of $\Psi$ and $\Upsilon$ mesons via  $p-p$  collisions has
been of interest for many years as a test of QCD (Quantum Chromodynamics). 
More than a decade ago it was shown that the relative production of
$\Psi(2S)$ to  $J/\Psi(1S)$ in $p-\bar{p}$ collisions was not consistent with 
standard QCD models\cite{CDF00}. Similarly, in experiments on $\Upsilon(nS)$ 
production via $p-p$ collisions it was found\cite{CMS11,CMS13} that 
$\Upsilon(3S)$ to $\Upsilon(1S)$ production is also not consistent with 
standard QCD models. In a theoretical study of $\Psi$ and $\Upsilon$ 
production via $p-p$ or $p-\bar{p}$ collisions\cite{klm11} it was shown that 
the relative probabilities of $\Psi(2S)$ to  $J/\Psi(1S)$ and $\Upsilon(3S)$ 
to $\Upsilon(1S)$ are consistent with experiment if the $\Psi(2S)$ and
$\Upsilon(3S)$ are mixed heavy hybrids, discussed below. The fact that
$\Psi(2S)$ is a mixed charmonium hybrid meson and $\Upsilon(3S)$ is a mixed
bottomonium hybrid meson, while $J/\Psi(1S)$ and $\Upsilon(1S)$ are standard
charmonium and bottomonium mesons is the basis for the present work.

  Recent experiments using $d-Au$ collisions\cite{PHENIX13,PHENIX14} and $p-Pb$ 
collisions\cite{ALICE14,ALICEz14} have shown a strong suppression, $S_A$, of 
$\Psi(2S)$ relative to  $J/\Psi(1S)$. As stated in these articles, 
this suppression cannot be explained by current theoretical 
models\cite{kop11,sha13,alb13,arl13,mcg13}. 
In an earlier study of $J/\Psi$ production and absorption\cite{spi99} two
scenarios were used for charmonium production: 1. charmonium states are
produced with the $c\bar{c}$ a color octet ($|c\bar{c}(8)g>$); and 2. charmonium
states are produced as a color singlet $|c\bar{c}>$, which is the standard
model. 

In the present work on $\Psi(2S)$ suppression scenario 1. of Ref\cite{spi99} 
is used, as is discussed in Ref\cite{klm11}. We estimate $S_A$ for both 
$J/\Psi(1S)$ and $\Psi(2S)$ for $p-Pb$ collisions using the mixed heavy 
hybrid theory, and show that the ratio of $S_A$ for $\Psi(2S)$ to  $J/\Psi(1S)$
is consistent with experiments. 
CMS experiments have measured $\Upsilon$ states suppression in Pb-Pb 
collisions\cite{cms11,cms12}, and estimated the yields of 
$\Upsilon(3S)$/$\Upsilon(1S)$ relative to those in p-p collisions\cite{cms12}.
We estimate this ratio using our mixed hybrid theory.

\newpage
  Next we briefly discuss the method of QCD Sum Rules, and how this was
used to show that the $\Psi(2S)$ and $\Upsilon(3S)$ are mixed heavy hybrids,
defined in the next section.

\section{Mixed Heavy Hybrid States via QCD Sum Rules}

The starting point of the method of QCD sum rules\cite{sz79} for finding 
the mass of the state referred to as A is the correlator,
 
\beq
\label{PiA}
       \Pi^A(x) &=&  \langle | T[J_A(x) J_A(0)]|\rangle \; ,
\eeq
with $| \rangle$ the vacuum state and
the current $J_A(x)$ creating the states with quantum numbers A.
The QCD sum rule is obtained  by equating a dispersion relation of $\Pi^A$
in momentum space to an operator product expansion of $\Pi^A$ using QCD
diagrams with quarks and gluons. After taking a Borel transform\cite{sz79}, 
${\mathcal B}$, in which the momentum variable is  replaced by the Borel mass, 
$M_B$, the QCD sum rule 
has the form
\beq
\label{QCDsumrule}
    && \frac{1}{\pi} e^{-M_A^2/M_B^2}
+ {\cal B} \int_{s_o}^\infty \frac{Im[\Pi_A(s)]}{\pi(s-q^2)} ds
     = {\cal B} \sum_k c_k^A(q) <0|{\cal O}_k|0> \; ,
\eeq
where $M_A$ is the lowest mass of a state with the properties of $A$ and
the right-hand side is the Borel transform of the operator product expansion
of $\Pi^A$. The operator that produces the mixed charmonium and hybrid 
charmonium states, with $b$ determined from the Sum Rule, is
\beq
\label{11}
        J_{C-HC} &=& b J_H + \sqrt{1-b^2} J_{HH} \; ,
\eeq
with $J_H|0> = |c\bar{c}(0)>, J_{HH}|0> = |[c\bar{c}(8)g](0)>$,
where $|c\bar{c}(0)>$ is a standard Charmonium state, while a hybrid 
Charmonium state $|[c\bar{c}(8)g](0)>$ has $c\bar{c}(8)$ with color=8 and a 
gluon with color=8. For the mixed hybrid Charmonium state produced
by $J_{C-HC}$  mass $M_A$ of Eq(\ref{QCDsumrule}) is called $M_{C-HC}$. 
 To find the mass $M_{C-HC}$ one plots the value of $M_{C-HC}^2$ vs $M_B^2$
using Eq(\ref{QCDsumrule}) with the quantities derived in Ref.\cite{lsk09}.
The solution for $M_{C-HC}$ is given by the minimum in the plot. Note that 
$M_{C-HC}\simeq M_B^2$ for a solution satisfying the method of QCD Sum Rules. 
This plot is shown in the figure below for (Eq(\ref{11})) $b^2=0.5$.
\vspace{1.4cm}

\begin{figure}[ht]
\begin{center}
\epsfig{file=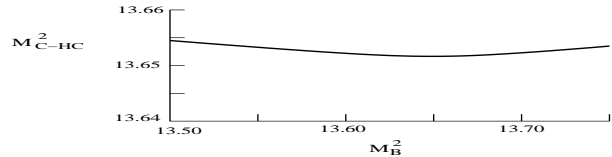,height=2cm,width=8cm}
\caption{Mixed Charmonium-hybrid charmonium mass $\simeq$ 3.65 GeV}
%\label{}
\end{center} 
\end{figure}
\newpage

  From this figure one sees that the minimum in $M^2_{C-HC}(M_B^2)$ corresponds
to the  $\Psi'(2S)$ state, with a mass\cite{PDG12} of 3.686 GeV. Therefore
the $\Psi'(2S)$ meson is 50\% normal Charmonium and 50\% hybrid 
Charmonium, while the $J/\Psi(1S)$ is a normal 
Charmonium meson. The analysis for Upsilon states was similar, with the 
$\Upsilon(3S)$ being 50\% normal Bottomonium  and 50\% hybrid Bottomonium, 
while the $\Upsilon(1S)$  and $\Upsilon(2S)$ states are standard Bottomonium 
mesons. We shall use this to estimate the ratio of suppression of  $\Psi(2S)$ 
to $J/\Psi(1S)$ in p-Pb collisions and $\Upsilon(3S)$ to $\Upsilon(3S)$ in
Pb-Pb collisions.

\section{Nuclear Modification and Suppression of 
$\Psi(2S)$/ $J/\Psi(1S)$ in p-Pb Cossisions}

In this section we derive the relative suppression of $\Psi(2S)$ to 
$J/\Psi(1S)$ and compare this result to experiment. First the definition of 
nuclear suppression and experimental data for the relative  $\Psi(2S)$ to 
$J/\Psi(1S)$ suppression is given, and then the theoretical derivation and 
comparison to experiment is presented.

 The mixed Charmonium hybrid theory, with the $\Psi'(2S)$ meson being 50\% 
normal Charmonium and 50\% hybrid Charmonium is directly used in calculating
the relative suppression.

\subsection{Experimental $\Psi(2S)$ to  $J/\Psi(1S)$ suppression in
p-Pb collisions}

 The nuclear modification for $\Phi=J/Psi(1S) {\rm \;or\;}\Psi(2S)$ produced
in A-B collisions is defined as\cite{PHENIX13,ALICE14}
\beq
\label{modification}
        R^\Phi &=& \frac{dN_\Phi^{A-B}/dy}{N_{coll} dN_\Phi^{pp}/dy} \; ,
\eeq
where $dN_\Phi^{A-B}/dy$ and $dN_\Phi^{pp}/dy$ are the invariant yields of 
$\Phi$ in A-B and pp collisions. In this work we consider p-Pb collisions 
(A=p, B=Pb).

The relative suppression of  $\Psi(2S)$ to  $J/\Psi(1S)$  is defined as
\beq
\label{suppression}
     R^{\Psi(2S)-J/\Psi(1S)}&=& \frac{R^{\Psi(2S)}}{R^{J/\Psi(1S)}} \; .
\eeq
The experimental resuls for  rapidity $0\leq y \leq 3$, as shown in the
figure below is
\beq
\label{exp-suppression}
        R^{\Psi(2S)-J/\Psi(1S)}|_{exp}&\simeq& 0.65 \pm 0.1
\eeq  

  As stated in Refs.\cite{PHENIX13,PHENIX14,ALICE14,ALICEz14}, the
observed suppression of $\Psi(2S)$ compared to $J/\Psi(1S)$ cannot be
explained in standard charmonium models. As stated by J. Matthew 
Durham\cite{PHENIX14}, ``the difference in suppression is too strong to
be explained by breakup effects in the nucleus...these observations raise
interesting questions about the mechanism of $\Psi(2S)$ suppression when
it is produced in a nuclear target.'' 

  Recently there was an attempt to explain the $\Psi(2S)$ versus $J/\Psi(1S)$ 
suppression using a comover interaction approach\cite{fer14}. In the present
work we show that the mixed hybrid theory for the $\Psi(2S)$ state, which
has been successful in predicting ratios of  $\Psi(2S)$ to $J/\Psi(1S)$
production cross sections in p-p\cite{klm11} and A-A\cite{klm14} collisions,
can explain the mystery of the he $\Psi(2S)$ versus $J/\Psi(1S)$ 
suppression.

\clearpage

The experimental results for  p-Pb (ALICE) and d-AU (PHENIX) collisions are 
shown in Figure 2, with $\sqrt{s_{NN}}=E_{NN}$ =nucleon-nucleon center of mass
energy.

\begin{figure}[ht]
\begin{center}
\epsfig{file=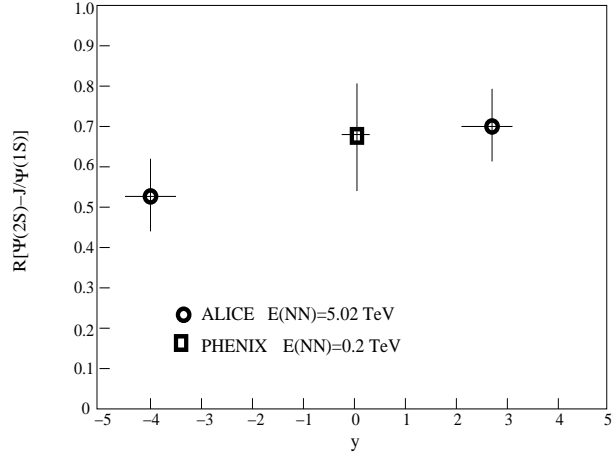,height=6cm,width=8cm}
\end{center}
\caption{The relative suppression of $\Psi(2S)$ to $J/\Psi(1S)$ for 
$E_{NN}$ =5.02 TeV p-Pb (ALICE) with rapidity $\simeq$ -4 and 
3; and $E_{NN}$=200 GeV d-Au (PHENIX) with rapidity $\simeq$ 0}
\label{}
\end{figure}
 
\subsection{Theoretical $\Psi(2S)$ to  $J/\Psi(1S)$ suppression in p-Pb
collisions}

The suppression, $S_A$, of charmonium states is given by the interaction with
nucleons as it traverses the nucleus. For a standard charmonium meson
state $|c\bar{c}>$ or hybrid meson state $|c\bar{c}g>$, with the $c\bar{c}$
having octet color, the equation for suppression is given by\cite{ks96} 
\beq
\label{SA}
    S_A &=& e^{-n_o\sigma_{\Phi N} L} \; ,
\eeq
where $\Phi$ is a $c\bar{c}$ or $c\bar{c}g$ meson,  $L$ is the length of 
the path of $\Phi$ in nuclear matter $\simeq$ 8 to 10 fm for
p-Pb collisions, with nuclear matter density 
$n_o=.017 fm^{-3}$, and $\sigma_{\Phi N}$ is the cross section for $\Phi$-
nucleon collisions.

 The cross section for standard charmonium $c\bar{c}$ meson via strong
QCD interactions with nucleons is given by\cite{ks96}
\beq
\label{sigma-c-barc} 
   \sigma_{c \bar{c} N} &=& 2.4 \alpha_s \pi r_{c \bar{c}}^2 \; ,
\eeq
where the strong coupling constant $\alpha_s\simeq $ 0.118\cite{PDG12}, and 
the charmonium meson radius $r_{c \bar{c}} \simeq \not h/(2 M_c c)$, with $M_c$ 
the charm quark mass. Using $2 M_c \simeq M_{J/\Psi} \simeq $ 3 GeV,
\beq
\label{rcc}
       r_{c \bar{c}} & \simeq&  \not h/(3 GeV c) \simeq 6 \times 10^{-17}m
= 0.06 fm
\eeq
  From Eqs(\ref{sigma-c-barc},\ref{rcc})
\beq
\label{sigmacbarc}
        \sigma_{c \bar{c} N} &\simeq & 3.2 \times 10^{-3} fm^2=
3.2 \times 10^{-2} mb \; .
\eeq

  Taking $L \simeq$ 8-10 fm and $n_o=.017 fm^{-3}$, from Eq(\ref{sigmacbarc}),
\beq
\label{SAcc}
        n_o\sigma_{c \bar{c} N} L &\simeq& 0.0022 \nonumber \\
        S_A^{c\bar{c}} &=& e^{- n_o\sigma_{c \bar{c} N} L} \simeq 1.0 \; .
\eeq

  On the other hand, the cross section for hybrid charmonium $c\bar{c}g$ meson 
via strong QCD interactions with nucleons has been estimated in Ref\cite{ks96}
as $\sigma_{c\bar{c}g N} \simeq$ 6-7 mb. In the present work we use
\beq
\label{sigmacbarcg}
     \sigma_{c\bar{c}g N} &\simeq & 6.5 mb \; .
\eeq

From this, using $L \simeq$ 8-10 fm and $n_o=.017 fm^{-3}$, from Eq(\ref{SA})
we obtain
\beq
\label{SAccg}
      n_o\sigma_{c\bar{c}g N} L&\simeq& 0.88 {\rm \;to\;}1.1 \nonumber \\
   S_A^{c\bar{c}g} &\simeq& 0.4 {\rm \;to\;} 0.33  \; .
\eeq

Using our mixed hybrid model, with 50\% $|c\bar{c}>$ and 50\% $|c\bar{c}g>$,
from Eqs(\ref{modification},\ref{SAcc},\ref{SAccg}), we find
\beq
\label{SAratio}
    R^{\Psi(2S)-J/\Psi(1S)}|_{theory}&\simeq& \frac{1+0.4{\rm \;to\;} 0.33}{2} 
\nonumber \\
          &=& 0.7{\rm \;to\;}0.66 \; .
\eeq  
Comparing Eqn(\ref{SAratio}) to Eqn(\ref{exp-suppression}), one finds that the
mixed hybrid theory for the state $\Psi(2S)$ solves the mystery of the  
large suppression of $\Psi(2S)$ vs $J/Psi$ in p-Pb collisions, and therefore
in other A-B collisions.

\section{Nuclear Modification and Suppression of  
$\Upsilon(3S)$/$\Upsilon(1S)$ in Pb-Pb Collisions}

This section is similar to the previous one, with the main difference being
that we use the experimental results of Ref\cite{cms12} for the ratios
of the standard $\Upsilon(3S)$ to $\Upsilon(1S)$ rather than the theoretical
estimate for $\Psi(2S)$ to $\Psi(1S))$ used in the previous section.

\subsection{Experimental $\Upsilon(3S)$ to  $\Upsilon(1S)$ 
suppression in Pb-Pb collisions}

As stated in Ref\cite{cms12}, although the ratios of observed yields
of $[\Upsilon(2S)/\Upsilon(1S)]_{pp}$, $[\Upsilon(2S)/\Upsilon(1S)]_{PbPb}$,
$[\Upsilon(3S)/\Upsilon(1S)]_{pp}$, and $[\Upsilon(3S)/\Upsilon(1S)]_{PbPb}$
must be corrected  for difference in acceptance and efficiency of the
$\Upsilon(2S)$ and $\Upsilon(3S)$ states to the $\Upsilon(1S)$ state,
by taking ratio of ratios these corrections are not needed. 

   The results for the ratio of ratios needed for the present work is
\beq
\label{3S1SPbpp}
  \frac{\Upsilon(3S)/\Upsilon(1S)|_{PbPb}}{\Upsilon(3S)/\Upsilon(1S)|_{pp}}
&=& 0.06 \pm 0.06 ({\rm stat})\pm 0.06 ({\rm syst}) ; .
\eeq

   We also use the result from Ref\cite{cms12} for the $\Upsilon(2S)$:
\beq
\label{2S1SPbpp}
  \frac{\Upsilon(2S)/\Upsilon(1S)|_{PbPb}}{\Upsilon(2S)/\Upsilon(1S)|_{pp}}
&=& 0.21 \pm 0.07 ({\rm stat})\pm 0.02 ({\rm syst}) ; .
\eeq

Since the $\Upsilon(2S)$ state in the theory of Ref\cite{lsk09}, upon which
the present work is based, is a standard $b\bar{b}$ state, we shall use
this modified by the relative bottomium to charmonium nucleation time\cite{ks96}
to estimate the suppression ratio for the standard component of the 
$\Upsilon(3S)$ state in the next subsection.

\subsection{Theoretical $\Upsilon(3S)$ to  $\Upsilon(1S)$ 
suppression in Pb-Pb collisions}

  In deriving $S_A^{c\bar{c}}$, the suppression for a standard model $c\bar{c}$
state we used Eq(\ref{sigma-c-barc}) to obtain the cross section for
standard charmonium-nucleon cross section. Since the $\Upsilon(2S)$ is
a standard $b\bar{b}$ state, we can get a more accurate result for standard
bottomium supression Pb-Pb to pp for the $b\bar{b}$ component of the
$\Upsilon(3S)$ from Eq({\ref{2S1SPbpp}) modified by the relative neutralization
time\cite{ks96} of $b\bar{b}$ vs $c\bar{c}$=$\sqrt{M_c/M_b}\simeq 0.55$
\beq
\label{upsilonstandard}
 S_A^{b\bar{b}}&=&\frac{\Upsilon(3S)/\Upsilon(1S)|_{PbPb}}
{\Upsilon(3S)/\Upsilon(1S)|_{pp}}|_{sm} \simeq 0.11 \; .
\eeq

  For the cross section for hybrid bottomonium $b\bar{b}g$ meson 
via strong QCD interactions with nucleons\cite{ks96} $\sigma_{b\bar{b}g N}=
\sigma_{c\bar{c}g N}(M_c/M_b)^2\simeq 0.09 \sigma_{c\bar{c}g N}$, therefore from 
Eq(\ref{sigmacbarcg}) 
\beq
\label{sigmabbarbg}
     \sigma_{b\bar{b}g N} &\simeq & 0.59 mb \; .
\eeq

Using $L \simeq$ 15 fm for Pb-Pb collisions and $n_o=.017 fm^{-3}$, 
from Eq(\ref{SA})
we obtain
\beq
\label{SAccg}
      n_o\sigma_{b\bar{b}g N} L&\simeq& 0.15 \nonumber \\
   S_A^{c\bar{c}g} &\simeq& 0.017  \; .
\eeq

From Eqs(\ref{upsilonstandard},\ref{SAccg}) one obtains
\beq
\label{SAbbratio}
    R^{\Upsilon(3S)-\Upsilon(1S)}|_{theory}&\simeq& \frac{.11 +.017}{2} 
\nonumber \\
          &\simeq& 0.06 \;,
\eeq  
in agreement with  the experimental ratio shown in Eq(\ref{3S1SPbpp}),
within experimental and theoretical errors.

\section{Conclusions}

Using our mixed hybrid theory for the $\Psi(2S)$ and $\Upsilon(3S)$
states we have found approximate agreement with experiment  for the $\Psi(2S)$ 
to $\Psi(1S)$ cross section ratio for p-Pb vs p-p collisions, and the 
$\Upsilon(3S)$ to $\Upsilon(1S)$ cross section ratio for Pb-Pb vs p-p
collisions.
\vspace{5mm}
\newpage

\Large
{\bf Acknowledgements}
\normalsize
\vspace{5mm}

The author thanks Dr. Debasish Das, Saha Institute of Nuclear Physics, for the
suggestion to consider the mixed hybrid heavy quark theory as an alternative 
to the theory used in Ref\cite{fer14} to explain the results in 
Refs\cite{PHENIX13,PHENIX14,ALICE14,ALICEz14}.

\end{document}